
\magnification=1200
\pretolerance=10000
\baselineskip=18pt
\voffset -1 true cm
\centerline {\bf STABLE DIQUARK MATTER ?}
\vskip 1 true cm
\centerline {J.E.Horvath}
\centerline {\it Instituto Astron\^omico e Geof\'\i sico}
\centerline {\it Universidade de S\~ao Paulo}
\centerline {\it Av. M. St\'efano 4200 - Agua Funda}
\centerline {\it (04301) S\~ao Paulo - SP - Brasil}
\vskip 3 true cm
\noindent
{\bf Abstract}

{\bf Two-quark correlations ({\it diquarks}) may play an important role in
 hadronic physics, particularly near the deconfinement point. This opens the
possibility of a net energy gain by means of a (non-perturbative) quark
pairing effect, perhaps up to stabilize diquark droplets. We address in
the present work the possibility of a self-bound, stable state of bulk diquark
matter.}
\vskip 8 true cm
\vfill\eject

\noindent
{\bf 1. Introduction}
\vskip 0.5 true cm
Considerable interest has been devoted to the physics of dense hadronic matter
in the last years. In addition to the astrophysical and cosmological
environments where a key role by QCD physics is expected, there is an obvious
motivation coming from the significative improvements in accelerator facilities
which may provide direct evidence of the state of matter above the saturation
density $\rho_{o}$. Of course we have not been able to solve the dynamics of
quarks and gluons in the non-perturbative regime and thus our knowledge of
very important issues such as the ground state energy and the onset of phase
transitions (chiral, deconfinement) remain uncertain. Lacking of reliable
computations in this regime, several phenomenological models have been devised
to address these points, the most popular one being the so-called M.I.T.
bag [1], which is considerably successful in reproducing most features of the
low-energy hadrons.

A particularly interesting suggestion related to these topics was the idea that
a high-strangeness variant of the quark-gluon plasma may be absolutely stable
[2] and thus the true ground state of hadronic matter. Witten's
{\it strange matter} has been the subject of activity concerning its various
properties and production/detection mechanisms, as well as the corresponding
astrophysical and cosmological consequences [3] (see also Ref.[4] for a
discussion of "strange baryon matter" catalyzed by kaon condensation).
 Amusingly, there have been also
suggestions of metastable or stable exotic particles made out of a few quarks.
 Examples of this class where {\it symmetry} properties are crucial for the
lifetime of the state are the {\it H} dihyperon [5] and the $Q_{\alpha}$ boson
[6].

Recently, another work [7] raised the possibility of substantial quark-quark
"hyperfine" interactions surviving above the deconfinement point [8].
Therefore,
in this {\it diquark} picture the quarks remain correlated and lower their
energy in this regime, being an intermediate state before the asymptotic
freedom region. We shall discuss in this work the possibility of an absolutely
stable diquark phase (quite analogous to the strange matter argument),
restricting ourselves to the region of high density and low temperatures, and
point out some phenomenological consequences to be analyzed elsewhere.
\vskip 0.5 true cm
\noindent
{\bf 2. Diquark physics}
\vskip 0.5 true cm
The diquark suggestion is based mainly on the assumption that the most relevant
interaction between quarks is of the form

$$ H_{I} = - A {\sum \atop {i \not= j}} b^{\dagger}_{i} \; \sigma^{a} \;
\lambda^{B} \; b_{i} \; b^{\dagger}_{j} \; \sigma^{a} \; \lambda^{B} \;
b_{j} , \eqno(1) $$

where $i,j$ label the quarks, $b, b^{\dagger}$ are their annhilation and
creation operators, $\sigma^{a}$ are the usual Pauli matrices and $\lambda^{B}$
the color $SU(3)$ ones. The coupling $A$ is univocally related to the strong
interactions coupling constant $\alpha_{S}$, but has been instead fixed by
fitting the $N-\Delta$ mass difference. In terms of the color-spin
wavefunctions [9] it can be checked that the most attractive channel is
$\mid \overline{3}, 0>$ for which

$$ <\overline{3},0 \mid H_{I} \mid \overline{3},0> = - \; 16 A \eqno(2) $$

Because of this fact, we shall refer to the color triplet, spin-zero , isospin
zero combination as the diquark [7,8].

A mass of the diquark of $m_{D} \simeq $ 575 MeV has been derived by assuming
that the $N-\Delta$ mass difference of about 300 MeV is determined by $H_{I}$.
 Hence, the expectation is that in the deconfined phase the third quark of a
nucleon should pair up with another free one to maximize the attractive
energy. As energies grow higher an increasing fraction of "ionized" diquarks
would be found, but the former may be totally dominant in a certain density
range in the case of self-gravitating matter [10,11]. The question to be
answered is if such a diquark matter can exist as a self-bound state, that is
to say if the energy gain is enough to stabilize a bit of that matter,
or is preferred only under external pressure.
\vskip 0.5 true cm
\noindent
{\bf 3. The stability of diquark matter}
\vskip 0.5 true cm
We shall address in this section the stability of a diquark collection. For the
same reasons than in the strange matter case, we do not expect that the region
of high-$T$, low-$\rho$ can be relevant to this issue because the entropic
contribution $-TS$ in the free energy generally desestabilizes non-topological
solitons of this kind with respect to an ordinary nucleon gas [12].
We are thus led to explore
the opposite case of low-$T$ and high-$\rho$, where diquarks have been
modelled by an effective Lagrangian for a color-triplet field $\phi$

$$ L = {1 \over 2} ( \partial_{\mu} \phi^{\dagger} \partial^{\mu} \phi \; - \;
m_{D} \phi^{\dagger} \phi) \; - \; \lambda {(\phi^{\dagger} \phi)}^{2} ,
\eqno(3) $$

(note that the coupling $\lambda$ differs by a factor of 4 from the usual
field-theoretical value). By employing a variant of the $P$-matrix formalism
of Jaffe and Low [13], Donoghue and Sateesh [7]
obtained the value $\lambda = 27.8$
which is indicative of the strenght of the repulsion that avoids a Bose
condensate at finite density. However, there is still nothing
in this description that tells of the fact that diquarks are {\it colored}
objects and therefore can not escape from the region where the vacuum is able
to support them. These confining interactions should probably come from the
neglected higher-order terms $ {\sum \atop {n>2}} \lambda_{n}
{(\phi^{\dagger} \phi)}^{n} $ in the Lagrangian of eq.(3) which are untractable
at present for our problem (note that as a
matter of fact the {\it mass} is also renormalized
by high-order interactions and should be properly denoted as $m_{D}^{\ast}$ as
in nuclear matter calculations, see below). We shall therefore proceed to
introduce a vacuum energy density term $\varepsilon_{V}$ analogous to the
M.I.T.
 bag constant $B$ to simulate confinement of diquarks in a finite volume
(of course any another confinement model may
be also adopted). It is known that in the latter case
a good fit to the hadronic spectroscopy can be obtained with a value
$B^{1/4} \simeq /; 145 $MeV [1]. In the diquark approach, this needs
not to be true
 and we should recalculate everything to extract a sensible value for
$\varepsilon_{V}$; whose value does not in principle
affect the determination of
$\lambda$, but is entangled with the derivation of $m_{D}$. We shall treat
it as a free parameter in the following of this work.

An approximate equation of state valid for self-inteacting bosons as described
by the Lagrangian of eq.(3) has been derived in Refs.[7,10] and will be adopted
here. That description consists of assuming a Gaussian distribution function
$f(k)$ for the diquarks

$$ f(k) = {N \over {2 \; \pi \; \sigma^{2}}^{3/2}}
exp({-k^{2} \over {2 \sigma^{2}}}) \eqno(4) $$

and minimize the energy derived from eq.(3) with respect to the width $\sigma$.
 In the limits of high  and low diquark densities $n_{D}$ the minimization can
be carried out analytically. Furthermore, it can be checked (see for example
the Fig.1 of Ref.[10]) that the low-density limit applies whenever
$n_{D} \leq 10 n_{o}$ and $f(k) \rightarrow \delta (k)$. This is safely within
the range we are interested in and justifies the adoption of

$$ P_{D} = {\lambda \over {2 m_{D}^{2}}} \; n_{D}^{2} \eqno(5a) $$
$$ \rho_{D} = m_{D} \; n_{D} \eqno(5b) $$

(where $P_{D} , \; \rho_{D}$ are the diquark pressure and energy density
respectively) for the equation of state.

We are now in position to investigate the possibility of a bound state of
diquarks. Depending on the flavor content
of the mixture, namely isoscalar matter or charge-zero matter, two cases
are possible and they will  be discussed separately.

\noindent
{\it a) Isoscalar matter}

Isoscalar matter having equal numbers of $u$ and $d$ quarks would form
($n_{u} \; + \; n_{d}$)/2 diquarks below a density
$\rho_{\ast} \simeq \; \rho_{o}$ [10] due to the lower value of $m_{D}$
compared to twice the value of the constituent quark mass.
Since the densities relevant for
diquark matter stability never exceed $\rho_{o}$ our assumption of the absence
of free quarks in this case (full pairing) will be
justified {\it a posteriori}.

Bulk matter must be electrically neutral and thus (relativistic) electrons
need to be present to satisfy this condition.
The electron thermodynamic quantities can be
derived from the gran canonical potential $\Omega = - \mu^{4}_{e}/12 \pi^{2}$
where $\mu_{e}$ is the electron chemical potential. In addition, the vacuum
effect $\varepsilon_{V}$ must be added to the energy and substracted
from the pressure as usually done. The program to compute
the existence of stable states is quite simple : first we impose the necessary
condition of a zero-pressure point

$$ P = P_{D} + P_{e} - \varepsilon_{V} = 0  \eqno(5) $$

which, together with the electrical neutrality condition

$$ {1 \over {3}} n_{D} - n_{e} = 0 \eqno(6) $$

is used to determine the maximum value of $\varepsilon_{V}^{1/4}$ which
still renders
a baryochemical potential
$\mu_{B} = (\sum \rho_{i} + P_{i})/ n_{B}$ lying below the nucleon mass $m_{n}$
for a given value of $m_{D}$
(strictly speaking, we should demand $\mu_{B}$ to be less than the energy
per baryon of a $ ^{56}Fe$ crystal, which is $\simeq 8$ MeV below $m_{n}$
but we have not taken into account this refinement). Because we expect that
the diquark mass $m_{D}$ gets non-negligible corrections from higher-order
contributions, the maximum $\varepsilon_{V}$ has been determined for a range
starting at $m_{D}^{\ast} = 0.8 m_{D}$ (which is a typical effective value
encountered for $m_{n}^{\ast}$ in many-body calculations) and going to the
highest value of $m_{D}$ for which stability
is conceivable. The resulting region is given
in Fig.1.

\noindent
{\it b) Charge-zero matter}

In this case the mixture already contains as many $d$ quarks as diquarks
(i.e. it can be thought as being made of partially "broken" neutrons) and no
electrons are needed to neutralize the bulk matter. We have assumed
non-relativistic free constituent quarks with mass $m_{d} = 360$ MeV. Thus,
eqs.(5) and (6) are replaced by

$$ P = P_{D} + P_{d} - \varepsilon_{V} = 0 \eqno(7) $$

$$ n_{D} - n_{d} = 0 \eqno(8) $$

respectively. The procedure is analogous to the isoscalar case, namely to
search for the values of $\varepsilon_{V}^{1/4}$ which make
$\mu_{B} \leq m_{n}$ as a function of $m_{D}$. The results are displayed
in Fig.2.

In both considered cases we have found that the maximum allowed values of
the vacuum energy density $\varepsilon_{V}^{1/4}$ are lower than the
"canonical" $B^{1/4} \simeq 145$ MeV obtained for the M.I.T. bag. In fact,
this is not very surprising since the naive sum of diquark + constituent
quark energies inside an hadron is already quite close to $m_{n}$ and
therefore such a model would not require a large value of $\varepsilon_{V}$
to fit its mass. We conclude from these results that a self-bound diquark
matter is in principle possible.

The zero-pressure,
self-bound diquark state is found to be less dense than strange matter
having a density $\rho \simeq 4 B \simeq \; 4 \; 10^{14} \; g \; cm^{-3}$.
 The density of the hypothetical
diquark chunks is mainly determined by the value of
$\rho_{D}$ and may lie between
$ \simeq 10^{13} \; g \; cm^{-3}$ to $ 1.9 \; 10^{14} \; g \; cm^{-3}$,
which are the extreme limits derived from the windows in Figs. 1 and 2.

It is important to remind that
even if the vacuum energy density happens to fall inside the stability
window it
is clear that the latter
should be in any case a bulk effect. This is to
avoid the spontaneous conversion of nuclei into a diquark soup. In the strange
matter picture a minimum (threshold) baryon number $A_{min} \simeq 10-100$ is
postulated to exist to preclude the conversion, because if so a simultaneous
decay of $\sim A/3$ quarks producing a net strangeness must occur to achieve
an energy gain. In the diquark case a similar situation holds, although strange
quark production is not needed in principle. The suppression is related to
the low probability for an isolated quark of a given nucleon to overlap with
another one of a neighbour nucleon (with the right symmetry numbers) inside
a nucleus. A few-diquark configuration should be desestabilized
by their interactions and to achieve a net energy gain we should reach the
bulk limit. In other words {\it confinement} acts as a barrier
against the decay and stabilization would only be possible once matter is
compressed beyond $\rho_{o}$ and the quarks are free to
recombine as they wish (this has been previously noted in Ref.[7]).

On the other hand, it should be noted that we have not {\it demonstrated} that
interacting diquarks are stable under decays of the type
$\phi \rightarrow \; q \overline{q}$. If they are not, the diquark matter would
have a lifetime of a fraction of second [14] which is nevertheless
very long for particle physics
standards and would lead to observable consequences.
\vskip 0.5 true cm
\noindent
{\bf 4. Conclusions and discussion}
\vskip 0.5 true cm
 We have entertained the possibility of a stable phase of matter composed of
diquarks, the latter being quark pairs in the maximally attractive state of
the hyperfine interaction Hamiltonian (eq.(1)). This would require a negative
pressure produced by the QCD vacuum properties lower than the M.I.T. value
$B^{1/4} \simeq 145 $MeV, but as far as we know, not excluded by any
experimental fact. The stability window has been found to be slightly larger
for the isoscalar than for the charge-zero case.

 On phenomenological grounds diquark matter strongly resembles the extensively
addressed strange matter case. Note, however, that there is an important
difference connected with the bulk electric charge. Strange matter needs to be
neutralized by electrons because the strange quark is massive and these
effects lead to a low $Z/A$ value for a given nugget. On the other hand, it
has been found that diquark matter can be found in either a isoscalar or
charge-zero content, depending on the parent nuclear matter.
This means that electrons are not always necessary for ensuring the electrical
neutrality of the chunk. A study
analogous to the one on strangelet formation [15] would then be interesting to
address the probability of a "diquarklet" production in the central-rapidity
region of heavy ion colliders, which are near the isoscalar case. On the other
hand it is possible that pure diquark stars may exist, having a maximum mass
[16] of $M_{MAX} \simeq 3 {\bigl( {\lambda \over {27.8}} \bigr)}^{1/2} \;
{\bigl( {575 MeV \over {m_{D}}} \bigr)}^{2} \; M_{\odot}$, perhaps covered
with a normal matter crust [17].
Note that this contrasts with the diquark stars
of Refs.[10,11] where diquark matter was {\it assumed} to appear only under
pressure. Several previous works on the combustion of nuclear matter [18],
supernovae [19], etc. made for the strange matter case can be carried over for
the diquark case as well and will be the subject of future publications.
\vskip 1 true cm
\noindent
{\bf Acknowledgements}

I would like to acknowledge the financial support of the Conselho Nacional de
Desenvolvimento Cient\'\i fico e Tecnol\'ogico (CNPq) , Brazil.
\vfill\eject
\noindent
{\bf References}

\noindent
1] A.Chodos, R.L.Jaffe, K.Johnson, C.B.Thorn and V.Weisskopf, {\it Phys. Rev.D}
{\bf 9} (1974) 3471.

\noindent
2] E.Witten, {\it Phys. Rev.D}{\bf 30} (1984) 272.

\noindent
3]See the {\it Proceedings of the Workshop on Strange Quark Matter in Physics
and Astrophysics}, Eds. J.Madsen and P.Haensel
{\it Nuc. Phys. B Proc. Supp.}{\bf 24} (1991) and the
extensive list of references given therein.

\noindent
4] B.W.Lynn, A.E.Nelson and N.Tetradis, {\it Nuc. Phys. B}{\bf 345} (1990) 186.

\noindent
5] R.L.Jaffe, {\it Phys. Rev. Lett.}{\bf 38} (1977) 195.

\noindent
6] F.C.Michel, {\it Phys. Rev. Lett.}{\bf 60} (1988) 667.

\noindent
7] J.F.Donoghue and K.S.Sateesh, {\it Phys. Rev.D}{\bf 38} (1988) 360.

\noindent
8] K.S.Sateesh, {\it Phys. Rev.D}{\bf 45} (1992) 866.

\noindent
9] J.Patera and D.Sankoff, {\it Branching Rules for Representation of Simple
Lie Algebras} (Les Presses de L'Universite de Montreal, Montreal,
Qu\'ebec 1973).

\noindent
10] D.Kastor and J.Traschen, {\it Phys. Rev.D}{\bf 44} (1991) 3791.

\noindent
11] J.E.Horvath, J.A.de Freitas Pacheco and J.C.N.de Ara\'ujo, to appear in
{\it Phys. Rev.D}.

\noindent
12] J.Frieman, A.V.Olinto, M.Gleiser and C.Alcock, {\it Phys.Rev.D}{\bf 40}
(1989) 3421.

\noindent
13] R.L.Jaffe and F.E.Low, {\it Phys. Rev.D}{\bf 19} (1979) 2105.

\noindent
14] A.Cohen, S.Coleman, H.Georgi and A.Manohar, {\it Nuc. Phys.B}{\bf 272}
(1986) 301 ; B.W.Lynn, {\it Nuc. Phys.B}{\bf 321} (1989) 465.

\noindent
15] C.Grainer, P.Koch and H.St$\ddot o$cker, {\it Phys. Rev. Lett.}{\bf 58}
 (1987) 1825.

\noindent
16] M.Colpi, S.L.Shapiro and I.Wasserman, {\it Phys. Rev. Lett.}{\bf 57}
(1986) 2485.

\noindent
17] C.Alcock, E.Farhi and A.V.Olinto, {\it Astrophys. J.}{\bf 310} (1986) 261.

\noindent
18] A.V.Olinto, {\it Phys. Lett.B}{\bf 192} (1987) 71.

\noindent
19] O.G.Benvenuto, J.E.Horvath and H.Vucetich, {\it Int. Jour. Mod. Phys.A}
{\bf 6} (1991) 4769.

\vfill\eject

\noindent
{\bf Figure captions}

\vskip 1 true cm
\noindent
{\bf Figure 1}. The stability window of isoscalar
diquark matter (hatched) as a function
of the vacuum energy density $\varepsilon_{V}$ and the (effective) diquark
mass $m_{D}$ (see the text for details). Both axis are measured in $MeV$.

\vskip 0.5 true cm
\noindent
{\bf Figure 2}. The same as in Fig.1 for charge-zero diquark matter.
\bye